\documentclass[10pt,final,journal]{IEEEtran}

\usepackage{graphicx,bm,amsmath,marvosym,nopageno,mathtools,amssymb,color,url,environ,array,tabularx}
\usepackage[ruled,vlined,linesnumbered,resetcount]{algorithm2e}

\renewcommand{\v}[1]{\boldsymbol{\mathbf{#1}}} 
\newcommand{\opn}[1]{\operatorname{#1}} 
\newcommand{\supp}[1]{\operatorname{supp}\left(\v{#1}\right)} 
\newcommand{\card}[1]{\left|\mathbb{#1}\right|} 
\newcommand{\set}[1]{\mathbb{#1}} 
\newcommand{\norm}[2]{\lVert #1 \rVert_{#2}} 
\newcommand{\braces}[1]{\left\{#1\right\}} 
\newcommand{\unset}[2]{\underset{#2}{#1}} 
\newcommand{\tm}[0]{\times} 
\newcommand{\prt}[1]{\left(#1\right)} 
\newcommand{\sqb}[1]{\left[#1\right]} 
\newcommand{\expt}[1]{\mathbb{E}\left\{#1\right\}} 
\newcommand{\abs}[1]{\left\lvert #1 \right\rvert} 
\newcommand{\inset}[2]{\left\langle #1\right\rangle_{#2}}

\newcommand{\inc}[1]{\in \set{C}^{#1}} 
\newcommand{\inr}[1]{\in \set{R}^{#1}} 
\renewcommand{\det}[1]{\operatorname{det}\left(#1\right)}
\DeclarePairedDelimiter\ceil{\lceil}{\rceil}
\DeclarePairedDelimiter\floor{\lfloor}{\rfloor}
\renewcommand{\vec}[1]{\operatorname{vec}\left(#1\right)}

\NewEnviron{meq}{
    \begin{align}
        \begin{split}
            \BODY
        \end{split}
    \end{align}
    }

\begin{document}

\title{Study of List-Based OMP and an Enhanced Model for Direction Finding with Non-Uniform Arrays}

\author{Wesley S. Leite, \IEEEmembership{Student Member, IEEE} and Rodrigo C. de Lamare, \IEEEmembership{Senior Member, IEEE}

    \thanks{This work was supported by the Pontifical Catholic University of Rio de Janeiro (PUC-Rio).}

    \thanks{Wesley S. Leite, and Rodrigo C. de Lamare are both with Centre for Telecommunications Research (CETUC), PUC-Rio, Rio de Janeiro 22451-900, Brazil. Rodrigo C. de Lamare is also with the Department of Electronics, University of York, York, U.K (e-mail: wleite@ieee.org, delamare@cetuc.puc-rio.br).}
}

\maketitle

\begin{abstract}
    This paper proposes an enhanced coarray transformation model (EDCTM) and a mixed greedy maximum likelihood algorithm called List-Based Maximum Likelihood Orthogonal Matching Pursuit (LBML-OMP) for direction-of-arrival estimation with non-uniform linear arrays (NLAs). The proposed EDCTM approach obtains improved estimates when Khatri-Rao product-based models are used to generate difference coarrays under the assumption of uncorrelated sources. In the proposed LBML-OMP technique, for each iteration a set of candidates is generated based on the correlation-maximization between the dictionary and the residue vector. LBML-OMP then chooses the best candidate based on a reduced-complexity asymptotic maximum likelihood decision rule. Simulations show the improved results of EDCTM over existing approaches and that LBML-OMP outperforms existing sparse recovery algorithms as well as Spatial Smoothing Multiple Signal Classification with NLAs.
\end{abstract}


\IEEEpeerreviewmaketitle

\section{Introduction}

\IEEEPARstart{T}{here} has been, recently, a great deal of research in the field of sparse reconstruction with Compressive Sensing (CS) algorithms. These techniques have been used for direction-of-arrival (DOA) estimation with several advantages over classical \cite{VanTrees2002} and subspace techniques \cite{intadap,jio,ccmcg,jidf,jidf_echo,sjidf,jiols,jiomimo,jioel,jiolcmv,jiostap,djio,ccmmwf,barc,rcmb,locsme,locsme_iet,dce,okspme,lrcc,l1stap,wljio,wlccm,blindmimo,armo,rrmser,memd,sor,utv,jiodoa,jioalrd,dynovs} as they can cost-effectively perform this task in challenging scenarios with excellent performance \cite{Stoeckle2015,Malioutov2005}.

In \cite{Ma2009}, an approach for quasi-stationary signals was proposed to estimate more sources than sensors, relying on the vectorization operator to build-up a full-rank effective array manifold. This technique was modified to deal with stationary signals through the introduction of a subspace-based algorithm known as Spatial Smoothing Multiple Signal Classification (SS-MUSIC) \cite{Pal2010,Pal2010-2} that performs spatial smoothing on what we call difference coarray transformation model (DCTM) in this paper. The DCTM model assumes perfectly uncorrelated sources, but does not account for finite-sample effects introduced by the vectorization operation. In addition, a sparse formulation of DCTM was proposed to deal with non-uniform linear arrays (NLAs) \cite{Zhang2013}.

This formulation allows the use of most CS algorithms such as Orthogonal Matching Pursuit (OMP) \cite{Pati1993}, Iterative Hard Thresholding (IHT) \cite{Blumensath2009}, and Compressive Sampling Matching Pursuit (CoSaMP) \cite{Needell2009}, but suffers from the same finite sample-effect observed in the model developed in \cite{Pal2010}. More recently, algorithms that exploit the sparsity property of the DOA spectrum have been developed, e.g., Sparse Bayesian Learning (SBL) and Focused OMP (FOMP) \cite{Gerstoft2016,Dehghani2018}.

Regarding classical DOA estimation, an asymptotic maximum likelihood (ML) rule was developed in \cite{Jaffer1988}. Despite being optimal from an estimation viewpoint in the mean-square sense, it is based on a nonlinear optimization procedure that is unfeasible for most practical scenarios with moderate to large number of source signals, since it requires a multi-dimensional grid search to find the optimal ML solution.

The objectives of this work are: i) to increase the accuracy of CS-based coarray DOA estimation with acceptable complexity when estimating support sets from deterministic largely coherent dictionaries; ii) to devise a cost-effective DOA estimation technique; and iii) to devise a novel coarray model that accounts for finite-sample errors in scenarios with uncorrelated sources and allows a more detailed analysis of the errors induced by the coarray transformation.

We propose an enhanced DCTM model (EDCTM) and a mixed greedy ML rule called List-Based Maximum Likelihood OMP (LBML-OMP) for DOA estimation with NLAs. The proposed EDCTM approach obtains improved angle estimates and emphasizes the errors introduced by Khatri-Rao product-based models for coarray DOA estimation with uncorrelated sources. In the case of LBML-OMP, a list of candidates is generated based on the correlation-maximization between the dictionary and the residue vector. The best candidate support set is then selected employing a modified reduced-complexity version of the asymptotic ML DOA estimator \cite{Jaffer1988}. The intuition behind this is that a set of candidate atoms could result in a higher likelihood of selecting improved support set elements. Since we select one atom at a time, the set of candidates is reduced to only one element through an optimum rule that maximizes the likelihood. Simulations show the improved results of EDCTM over the standard coarray model and that LBML-OMP outperforms existing algorithms like OMP, IHT, CoSaMP, ROMP and SS-MUSIC.

\emph{Paper structure}: In Section \ref{sec:systemModelandProblemStatement}, the data model for the DOA estimation problem with the standard difference coarray formulation is described. In Section \ref{sec:CsCoarrayDoAEstimation}, the sparse version of this problem is presented so that it can be solved using CS, and the devised coarray model is described. In Section \ref{sec:PropLBMLOMP}, the proposed LBML-OMP is detailed. In Section \ref{sec:resultsAndDiscussion}, simulations are presented and discussed, whereas Section \ref{sec:conclusions} draws the conclusions.

\emph{Notation}: $\set{S}$, $a$, $\v{a}$ and $\v{A}$ indicate sets, scalars, column vectors, and matrices, respectively. $\v{P}_{\v{A}}=\v{A}(\v{A}^{H}\v{A})^{-1}\v{A}^{H}$ and $\v{P}_{\v{A}}^{\perp} = \v{I}-\v{P}_{\v{A}}$ are the projection matrices onto the range of $\v{A}$ and null-space of $\v{A}^{H}$. $\inset{\v{a}}{\set{L}}$ and $\inset{\v{A}}{\set{L}}$ are restrictions on the elements and the columns indexed by the set $\set{L}$, respectively. For example, if $\set{L}=\{3,4,1\}$ and $\v{a}=\sqb{5,-7,0, 6}^\top$, then $\inset{\v{a}}{\set{L}}=\sqb{0,6,5}^\top\inc{\card{L}}$. $\v{A}\circ \v{B}$ and $\v{A}\otimes \v{B}$ indicate the Khatri-Rao and Kronecker products, respectively. $\set{S}\setminus \set{L}$ denotes the difference between the sets $\set{S}$ and $\set{L}$. $\abs{\set{D}}$ is the cardinality of the set $\set{D}$. $\sqb{N}$ and $[\v{a}]_i$ means the set $\braces{1,\ldots,N}$ and the $i$-th entry of $\v{a}$, respectively.

\section{System Model and Problem Statement}\label{sec:systemModelandProblemStatement}

Consider a linear array with $N$ sensors at normalized positions defined by the set of integers $\set{S}$, illuminated by $D$ narrowband spatially and temporally uncorrelated sources with DOAs $\v{\theta}\inr{D}$ taken in relation to the array broadside axis. The received signal covariance matrix is given by \cite{Liu2012-555,Liu2016}
\begin{equation}\label{eq:recSigCovMat}
    \v{C}_{\v{x}_{\set{S}}} = \v{A}_{\set{S}}\prt{\v{\theta}}\opn{diag}\sqb{\sigma^2_1,\ldots,\sigma^2_D}\v{A}_{\set{S}}^H\prt{\v{\theta}}+\sigma_n^2\v{I},
\end{equation}
where $\v{x}_{\set{S}}= \v{A}_{\set{S}}(\v{\theta}) \v{s}+\v{n}_{\set{S}} \inc{N}$ is the received signal, $\v{s} \inc{D}$ is the vector of source signals, $\v{n}_{\set{S}} \inc{N}$ is the noise vector, $\v{A}_{\set{S}}(\v{\theta})$ is the array manifold matrix,   $\v{C}_{\v{x}_{\set{S}}}\inc{N\tm N}$ is the received signal covariance matrix, $\v{C}_{\v{s}}=\opn{diag}\sqb{\sigma^2_1,\ldots,\sigma^2_D}\inc{D\tm D}$ is the uncorrelated zero-mean Gaussian sources covariance matrix, and $\v{C}_{\v{n}_{\set{S}}}=\sigma_n^2\v{I}\inc{N\tm N}$ represents the zero-mean additive white Gaussian noise covariance matrix. The source signals and the noise are assumed to be uncorrelated, and $\v{C}_{\v{x}_{\set{S}}}$ can be estimated via the sample covariance matrix estimate denoted by $\hat{\v{C}}_{\v{x}_{\set{S}}}$ \cite{Carlson1988}.

An approach to increasing the available degrees of freedom of the array is the difference coarray transformation using the $\vec{}$ operator in (\ref{eq:recSigCovMat}), which allows us to obtain \cite{Ma2010}
\begin{equation}\label{eq:vecRecSigCovMat}
    \vec{\v{C}_{\v{x}_{\set{S}}}} = \prt{\v{A}_{\set{S}}^{\ast}\prt{\v{\theta}}\circ\v{A}_{\set{S}}\prt{\v{\theta}}}\v{p}+\sigma_n^2\tilde{\v{i}},
\end{equation}
where $\v{p}=\sqb{\sigma^2_1,\ldots,\sigma^2_D}^{\top}$ contains the source signal powers and $\tilde{\v{i}}$ refers to the vectorization of $\v{I}$.

Define the vectors $\v{a}$ and $\v{c}$, such that $\sqb{\v{a}}_k =n_k,~\forall~n_k\in\set{S}$, $k\in\sqb{N}$, and $\v{c}=\v{a}\otimes\v{1}_N-\v{1}_N\otimes\v{a}$, where $\v{1}_N$ is the all-ones column vector of dimension $N$. Clearly, $\v{c}$ consists of all pairwise differences in the sensors normalized positions, with repetition among its entries. Suppose there is a set of indices $\set{H}$ such that the indexing operation $\v{c}^\prime = \inset{\v{c}}{\set{H}}$ removes the repeated elements in $\v{c}$ after their first occurrence and sort them in ascending order. The entries of $\v{c}^\prime\inr{\card{D}}$ define the set $\set{D}$
\begin{equation}
    \set{D}=\braces{n_i-n_k |\prt{n_i,n_k}\in \set{S}^2\text{, and } 1\leq i,k \leq N}
\end{equation}
whose cardinality $\card{D}$ is termed degrees of freedom (DoF). Notice that the set $\set{H}$ is directly determined from the array geometry $\set{S}$. Indexing both sides of (\ref{eq:vecRecSigCovMat}) through $\set{H}$, leads us to the DTCM model \cite{Pal2010}:
\begin{align}
    \begin{split}\label{eq:vecIndexedRecSigCovMat}
        \v{x}_{\set{D}} & = \sqb{\begin{array}{@{}c|c@{}}
                \v{A}_\set{D}(\v{\theta}) & \v{i} \\
            \end{array}}
        \sqb{
            \begin{array}{@{}c|c@{}}
                \v{p}^{\top} & \sigma_n^2 \\
            \end{array}
        }^{\top},
    \end{split}
\end{align}
where $\v{A}_{\set{D}}\prt{\v{\theta}}\inc{\textrm{DoF}\tm D}$ is the coarray manifold matrix (row-indexing of $\v{A}_{\set{S}}^{\ast}\prt{\v{\theta}}\circ\v{A}_{\set{S}}\prt{\v{\theta}}\inc{N^2\tm D}$ through $\set{H}$), $\v{x}_{\set{D}}\inc{\textrm{DoF}}$ is the coarray received signal, and $\sqb{\v{i}}_{k}=\delta(k-(\textrm{DoF}+1)/2)\in \braces{0,1}^{\textrm{DoF}}$, for $k\in \sqb{\textrm{DoF}}$.

\section{Proposed Enhanced Model and CS-Based DOA Estimation}\label{sec:CsCoarrayDoAEstimation}

Consider a grid $\v{\theta}^{g}\inr{g}$ that contains the true DOAs $\v{\theta}$. The dense DTCM in (\ref{eq:vecIndexedRecSigCovMat}) is cast with $\v{\theta}^{g}$ to include the $D$-sparse vector $\v{p}^g\inc{g}$ such that we obtain the sparse DCTM as \cite{Zhang2013}:
\begin{align}
    \begin{split}\label{eq:CSvecIndexedRecSigCovMat}
        \v{x}_{\set{D}} & = \sqb{\begin{array}{@{}c|c@{}}
                \v{A}_\set{D}(\v{\theta}^g) & \v{i} \\
            \end{array}}
        \sqb{
            \begin{array}{@{}c|c@{}}
                \v{p}^{g\top} & \sigma_n^2 \\
            \end{array}}^{\top}\\
        & = \v{B}\v{h},\text{ with } \norm{\v{h}}{0}\ll g+1,
    \end{split}
\end{align}
where $\v{B}\inc{\textrm{DoF}\tm (g+1)}$ is the dictionary, $\v{h}\inc{g+1}$ is a $K$-sparse vector ($K=D+1$), and $\v{x}_{\set{D}}$ is the measurement vector. We remark that the model in (\ref{eq:CSvecIndexedRecSigCovMat}) is the sparse version of the DCTM model in (\ref{eq:vecIndexedRecSigCovMat}).

\subsection{Proposed EDCTM model}\label{sec:edctm}

In order to account for scenarios with finite snapshots that arise naturally in DOA estimation, an alternative to (\ref{eq:vecIndexedRecSigCovMat}), termed Enhanced DCTM (EDCTM), has been devised as follows. From what has been previously explained, the model in (\ref{eq:recSigCovMat}) assumes spatially uncorrelated sources and noise. From that, $\v{C_{s}}$ is diagonal. This assumption makes the DCTM model simpler to compute. This is the formulation usually found in the literature \cite{Pal2010,Ma2010} and used to derive (\ref{eq:vecIndexedRecSigCovMat}). However, evaluating (\ref{eq:recSigCovMat}) under a finite-snapshot perspective, it follows that $\hat{\v{C}}_{{\v{x}_\set{S}}} \approx \v{A}_\set{S}(\v{\theta})\hat{\v{C}}_{\v{s}}\v{A}_{\set{S}}^H(\v{\theta})+\hat{\v{C}}_{\v{n}_\set{S}}$, where the sources-noise crossed terms were removed because we are mainly interested in the finite-sample effect included in $\hat{\v{C}}_{\v{s}}$ and DCTM also does not account for these terms. The $\vec{\cdot}$ operator property $\vec{\v{ABC}}=\prt{\v{C}^T\otimes \v{A}}\vec{\v{B}}$ allows us to write
\begin{align}\label{eq:difcobeforeindexing}
    \begin{split}
        \vec{\hat{\v{C}}_{{\v{x}_\set{S}}}} & = \prt{\v{A}_\set{S}^{\ast}(\v{\theta})\otimes \v{A}_\set{S}(\v{\theta})}\opn{vec}\left(\opn{diag}\prt{\hat{\v{C}}_{\v{s}}}+\hat{\v{C}}_{\v{s}}\right.\\
        & \left.-\opn{diag}\prt{\hat{\v{C}}_{\v{s}}}\right)+\vec{\hat{\v{C}}_{\v{n}_\set{S}}}\\
        & = \prt{\v{A}_\set{S}^{\ast}(\v{\theta})\otimes \v{A}_\set{S}(\v{\theta})}\vec{\opn{diag}\prt{\hat{\v{C}}_{\v{s}}}}\\
        & +\prt{\v{A}_\set{S}^{\ast}(\v{\theta})\otimes\v{A}_\set{S}(\v{\theta})}\vec{\hat{\underline{\v{C}}}_{\v{s}}}+\vec{\hat{\v{C}}_{\v{n}_\set{S}}}\\
        & = \prt{\v{A}_\set{S}^{\ast}(\v{\theta})\circ \v{A}_\set{S}(\v{\theta})}\opn{vecd}\prt{\hat{\v{C}}_{\v{s}}}\\
        &+\overbrace{\vec{\v{A}_\set{S}(\v{\theta})\hat{\underline{\v{C}}}_{\v{s}}\v{A}_{\set{S}}^H(\v{\theta})}}^{ \let\scriptstyle\textstyle\substack{\textrm{error term }\v{\eta}}}+\vec{\hat{\v{C}}_{\v{n}_\set{S}}},
    \end{split}
\end{align}
where the error term is a function of the hollow matrix $\hat{\underline{\v{C}}}_{\v{s}} = \hat{\v{C}}_{\v{s}}-\opn{diag}\prt{\hat{\v{C}}_{\v{s}}}$ and $\opn{vecd}(\cdot)$ is the linear diagonal extraction operator, which maps a matrix into a column vector with entries equal to the main diagonal elements, i.e., $\opn{vecd}\prt{\hat{\v{C}}_{\v{s}}}=\sum_{i=1}^{D}\v{P}_{\v{e}_i}\hat{\v{C}}_{\v{s}}\v{e}_i$, where $\v{P}_{\v{e}_i}=\v{e}_i\v{e}_i^T$ stands for the orthogonal projection operator onto the one-dimensional space spanned by the canonical basis vector $\v{e}_i$. From (\ref{eq:difcobeforeindexing}), it follows that
\begin{equation}\label{eq:etaeta}
    \vec{\hat{\v{C}}_{{\v{x}_\set{S}}}}-\v{\eta}=\prt{\v{A}_\set{S}^{\ast}(\v{\theta})\circ \v{A}_\set{S}(\v{\theta})}\hat{\v{p}}+\vec{\hat{\v{C}}_{\v{n}_\set{S}}}.
\end{equation}
Assuming $\inset{\vec{\hat{\v{C}}_{\v{n}_{\set{S}}}}}{\set{H}}\approx \hat{\sigma}_{n}^2 \v{i}$ and indexing both sides of (\ref{eq:etaeta}) through $\set{H}$, leads to the proposed dense EDCTM given by
\begin{equation}\label{eq:newDoA_DAQ_coarray}
    \hat{\v{x}}_{\set{D}}^{\rm EDCTM}=\hat{\v{x}}_\set{D}-\v{\eta}^{\prime} = \sqb{\begin{array}{@{}c|c@{}}
            \v{A}_\set{D}(\v{\theta}) & \v{i} \\
        \end{array}}
    \sqb{
        \begin{array}{@{}c|c@{}}
            \hat{\v{p}}^{\top} & \hat{\sigma}_n^2 \\
        \end{array}
    }^{\top},
\end{equation}
with corresponding sparse version assuming the form
\begin{meq}\label{eq:newDoA_DAQ_coarray_sparse}
    \hat{\v{x}}_{\set{D}}^{\rm EDCTM} & =\hat{\v{x}}_\set{D}-\v{\eta}^{\prime}\\
    & = \sqb{\begin{array}{@{}c|c@{}}
            \v{A}_\set{D}(\v{\theta}^{g}) & \v{i} \\
        \end{array}}\sqb{
        \begin{array}{@{}c|c@{}}
            \hat{\v{p}}^{g\top} & \hat{\sigma}_n^2 \\
        \end{array}
    }^{\top}\\
    & = \v{B}\hat{\v{h}},\text{ with } \norm{\hat{\v{h}}}{0}\ll g+1,
\end{meq}
where $\v{\eta}^{\prime}=\inset{\v{\eta}}{\set{H}}$. This model represents the acquired data with more precision and can be used for coarray DOA estimation by CS-based or conventional algorithms, under the assumption of previous knowledge of $\v{\eta}^{\prime}$ or the availability of its estimated version given by $\hat{\v{\eta}}^{\prime}_{\Delta}$, where $\Delta$ is an estimation procedure. Besides, it brings the attention to the fact that an important error is introduced when transforming standard arrays into coarrays for finite-snapshot scenarios with uncorrelated sources.

\subsection{Asymptotic properties and statistical behavior of EDCTM}\label{sec:asympprop}

The main goal of EDCTM is to remove the finite sample effect from the coarray model introduced by the sample covariance matrix of uncorrelated sources. The notion that this effect is more prominent for small-sample scenarios becomes clear through the analysis of the asymptotic behavior of $\v{\eta}^\prime$ (number of snapshots $T$ increases indefinitely). In order to verify that, we proceed as follows.

The sample covariance matrix follows a complex Wishart statistical distribution, since the sources are assumed to be zero-mean multivariate complex Gaussian distributed \cite{Goodman1963}, i.e., $\hat{\v{C}}_{\v{s}}\thicksim \mathcal{CW}_{D}(T,(1/T)\v{C_s})$, where D sources are considered, there are $T$ snapshots available (Complex Wishart degrees of freedom) and $\v{C}_{\v{s}}=\opn{diag}(\v{p})$ represents the source signals covariance matrix. The moments for the elements of $\hat{\v{C}}_{\v{s}}$ were obtained in \cite{Maiwald2000} through the differentiation of the characteristic function of the random variables $\sqb{\v{C}_{\v{s}}}_{11},\ldots,\sqb{\v{C}_{\v{s}}}_{D,D},2\mathfrak{Re}\braces{\sqb{\v{C}_{\v{s}}}_{12}},2\mathfrak{Im}\braces{\sqb{\v{C}_{\v{s}}}_{12}},\ldots,$ $2\mathfrak{Re}\braces{\sqb{\v{C}_{\v{s}}}_{N-1,N}},2\mathfrak{Im}\braces{\sqb{\v{C}_{\v{s}}}_{N-1,N}}$, resulting in
\begin{align}\label{eq:momentswishart}
    \begin{split}
        \expt{\sqb{\hat{\v{C}}_{\v{s}}}_{ij}} & =\sqb{\v{C_s}}_{ij},\\
    \end{split}
\end{align}
and
\begin{align}\label{eq:momentswishart2}
    \begin{split}
        \expt{\sqb{\hat{\v{C}}_{\v{s}}}_{ij}\sqb{\hat{\v{C}}_{\v{s}}}_{ji}} & = \expt{\sqb{\hat{\v{C}}_{\v{s}}}_{ij}\sqb{\hat{\v{C}}_{\v{s}}}_{ij}^{\ast}}\\
        & = \expt{\abs{\sqb{\hat{\v{C}}_{\v{s}}}_{ij}}^2}\\
        & =\abs{\sqb{\v{C}_{\v{s}}}_{ij}}^2+\frac{1}{T}\sqb{\v{C}_{\v{s}}}_{ii}\sqb{\v{C}_{\v{s}}}_{jj}.
    \end{split}
\end{align}

From this, since $\sqb{\v{C}_{\v{s}}}_{ij}=0\; \forall\; i\neq j$, then from (\ref{eq:momentswishart}) we can state that $\expt{\hat{\underline{\v{C}}}_{\v{s}}}=\v{0}$, which yields
\begin{align}\label{eq:expteta}
    \begin{split}
        \expt{\v{\eta}^\prime} & = \expt{\inset{\vec{\v{A}_\set{S}(\v{\theta})\hat{\underline{\v{C}}}_{\v{s}}\v{A}_{\set{S}}^H(\v{\theta})}}{\set{H}}}\\
        & = \inset{\vec{\v{A}_\set{S}(\v{\theta})\expt{\hat{\underline{\v{C}}}_{\v{s}}}\v{A}_{\set{S}}^H(\v{\theta})}}{\set{H}}\\
        & = \v{0}.
    \end{split}
\end{align}

Notice that the expected value of $\v{\eta}^{\prime}$ does not depend on the number of snapshots. Moreover, from (\ref{eq:momentswishart2}), the variance for the off-diagonal elements of $\hat{\underline{\v{C}}}_{\v{s}}$ is given by
\begin{equation}
    \expt{\abs{\sqb{\hat{\underline{\v{C}}}_{\v{s}}}_{ij}}^2}=\frac{1}{T}\sqb{\v{C}_{\v{s}}}_{ii}\sqb{\v{C}_{\v{s}}}_{jj}\; \forall\; i\neq j,
\end{equation}
which clearly states the inversely proportional dependence of this variance on the number of snapshots. Then, we have
\begin{equation}
    \lim_{T\to\infty}\expt{\abs{\sqb{\hat{\underline{\v{C}}}_{\v{s}}}_{ij}}^2}=0,
\end{equation}
i.e., the variance collapses to zero and the error term $\v{\eta}^{\prime}$ tends to not deviate from its expected value, i.e., the zero value. This explains why DCTM and EDCTM become the same model when enough data becomes available. This will be illustrated in Section \ref{sec:resultsAndDiscussion} through numerical simulations.

\emph{Remark}: Unlike the work of Koochakzadeh and Pal \cite{Koochakzadeh2016} in which the impact of off-diagonal terms of the source covariance matrix on the data vector is discussed, EDCTM is concerned with uncorrelated sources and deals with the more practical finite-sample effect of the source \emph{sample} covariance matrix (few snapshots), whereas the model (3) in \cite{Koochakzadeh2016} is intended to be used with correlated signals and was developed under the ideal infinite-snapshot condition. Therefore, the off-diagonal terms, and thus the correction terms, have different statistical characterization.

\section{Proposed LBML-OMP Algorithm}\label{sec:PropLBMLOMP}
The OMP algorithm is one of the most used sparse recovery strategies in CS \cite{Pati1993,Foucart2013,Aghababaiyan2020}. Its $i$-th iteration performs an atom search such that the maximum correlation between a given atom from the dictionary and the residual vector is found (correlation-maximization (corr-max)  step), according to
\begin{equation}\label{eq:ompSelectAtom}
    j= \unset{\opn{argmax}}{\hat{j}}\braces{\abs{\sqb{\v{F}^H\prt{\v{y}-\v{Fu}^{(i)}}}_{\hat{j}}}},
\end{equation}
where $\v{y}$ is the measurement vector, $\v{F}$ is the dictionary, and $\v{u}$ is the $K$-sparse vector to be found. Indeed, for scenarios with low SNR and few snapshots available, the OMP choice is not necessarily correct. The rationale is that, due to the nature of the deterministic dictionary, the neighbors of the OMP choice altogether increase the probability of containing the right solution.

The first step of LBML-OMP consists of a corr-max performed in the atoms indexed by $\set{J} = \sqb{g}\setminus \set{T}^{(i)}$, where $\set{T}^{(i)}$ is the support set from the previous iteration. This prevents an atom from being selected twice. Consider the index $n_0$ such that $j=\sqb{\set{J}}_{n_0}$, which is a restriction on $\set{J}$ to the element indexed by $n_0$. From that, the set with candidate indices is compactly represented by $ \set{V}=\braces{\sqb{\set{J}}_{n_0-n}~\biggl\lvert~ n=\floor*{-\frac{Q-1}{2}},\ldots,\ceil*{\frac{Q-1}{2}}}$, where $Q$ candidates are considered and $\floor*{\cdot}$ / $\ceil*{\cdot}$ are the \emph{floor} and \emph{ceil} functions, respectively. In simple terms, $\set{V}$ consists of the Q nearest neighbours on $\set{J}$ around $j$ (OMP index estimate), including $j$ itself.

The candidate support sets for $\v{p}^{g}$ are given by the union of each of the elements in $\set{V}$ with the support set from the previous iteration excluding the element $\braces{g+1}$, i.e., $\set{Y}_q=\braces{W_{N-1}\prt{\set{T}^{(i)}}\setminus \braces{g+1}}\cup v_q$, where $v_q\in \set{V}$, $q\in [Q]$ and the $W_{N-1}(\cdot)$ operator keeps only the $N-1$ latest elements added to a given set. After the support sets $\set{Y}_q$ are obtained, an optimum score rule must be devised to select the best possible approximation to the true DOAs $\v{\theta}$. A modified version of the asymptotic ML rule devised in \cite{VanTrees2002,Jaffer1988} is employed to perform this optimization procedure. Departing from zero-mean Gaussian complex sources, with i.i.d snapshots, the maximization of the log-likelihood function leads us to
\begin{equation}\label{eq:optprob}  \unset{\opn{maximize}}{\v{\theta},\v{C}_{\v{s}},\sigma_n^2}\quad-\sqb{\ln \det{\v{C}_{\v{x}_{\v{\set{S}}}}}+\opn{tr}\sqb{\v{C}_{\v{x}_{\v{\set{S}}}}^{-1}\hat{\v{C}}_{\v{x}_{\v{\set{S}}}}}},
\end{equation}
which has a separable solution for $\sigma_n^2$ and $\v{C}_{\v{s}}$. This solution can be used to obtain explicit functions of $\v{\theta}$. Replacing these functions into (\ref{eq:optprob}) we obtain the asymptotic (stochastic) ML estimator. After that, its D-dimensional domain is transformed into the domain built up by the candidates of the list. The rule for the optimum candidate selection becomes
\begin{equation}\label{eq:optProblbml-omp}
    \hat{q} \leftarrow \unset{\opn{argmin}}{q\in [Q]}\left\{\opn{det}\left(\v{U}_{q}\hat{\v{C}}_{\v{x}_{\set{S}}}\v{U}_{q}+\frac{\opn{tr}\prt{\v{U}_{q}^{\perp}\hat{\v{C}}_{\v{x}_{\set{S}}}}}{N-\lvert\set{Y}_q\rvert}\v{U}_{q}^{\perp}\right)\right\},
\end{equation}
where $\v{U}_{q}=\v{P}_{\inset{\v{A}_{\set{S}}\prt{\v{\theta}^g}}{\set{Y}_q}}$. From that, the $\hat{q}$-th element of $\set{V}$ is selected after a one-dimensional search on $Q$ points and its corresponding index $v_{\hat{q}}$ is added to the target support set. The $W_{N-1}(\cdot)$ operator plays the role of ensuring that the over the grid array manifold $\v{A}_{\set{S}}\prt{\v{\theta}^g}$ with atoms indexed by $\set{Y}_q$ is full-column rank, thus allowing the appropriate calculus of the projection matrix $\v{U}_q$ and LBML-OMP to estimate more sources than sensors.

The estimated K-sparse solution is obtained through the orthogonal projection of $\v{x}_{\set{D}}$ onto the spanning of the columns of $\v{B}$ indexed by the new target support set. This corresponds to the best possible $\set{T}^{(i+1)}$-sparse approximation to the solution of (\ref{eq:CSvecIndexedRecSigCovMat})  in the sense of the $l_2$-norm and can be obtained through
\begin{equation}
    \v{h}^{(i+1)}= \unset{\opn{argmin}}{\v{z}\inc{g+1}}\braces{\norm{\v{x}_{\set{D}}-\v{\v{Bz}}}{2}\textrm{, }\supp{z}\subset \set{T}^{(i+1)}},
\end{equation}
After $K$ iterations, the final DOA estimates are obtained according to $\hat{\v{\theta}}=\inset{\v{\theta}^g}{\set{T}^{(K)}\setminus \braces{g+1}}$. The LBML-OMP algorithm is summarized in Algorithm \ref{alg:lbml-omp}.

\begin{algorithm}[h!]
    \DontPrintSemicolon
    \SetKwInOut{Inp}{Input}
    \SetKwInOut{Out}{Output}

    \Inp{dictionary $\v{B}\inc{\textrm{DoF}\tm (g+1)}$, observed data $\v{x}_{\set{D}}\inc{DoF}$, sparsity level $K=D+1$, array manifold matrix over the grid $\v{A}_{\set{S}}\prt{\v{\theta}^g}\inc{N\tm g}$, sample received signal covariance matrix $\hat{\v{C}}_{\v{x}_{\set{S}}}\inc{N\tm N}$, number of sensors $N$, number of candidates $Q$, grid $\v{\theta}^g\inr{g}$}

    \BlankLine
    $\set{T}^{(0)}=\braces{g+1}$, $\v{h}^{(0)}=\v{0}$\label{alg:lbml-omp_initialize}
    \BlankLine

    \For{$i\leftarrow 0$ \KwTo $K-1$}{
    $\set{J}\leftarrow \sqb{g}\setminus \set{T}^{(i)}$\;\label{alg:lbml-omp_setSearch}
    $j^{(i)}\leftarrow\unset{\opn{argmax}}{\hat{j}\in\set{J}}\braces{\abs{\sqb{\v{B}^H\prt{\v{x}_{\set{D}}-\v{Bh}^{(i)}}}_{\hat{j}}}}$\;\label{alg:lbml-omp_corrMaxStep}
    $n_0\leftarrow $ position index of $j^{(i)}$ in $\set{J}$\;
    $\set{V}=\braces{\sqb{\set{J}}_{n_0-n}~\bigl\lvert~ n=\floor*{-\frac{Q-1}{2}},\ldots,\ceil*{\frac{Q-1}{2}}}$\;\label{alg:lbml-omp_chooseNearestNeighbours}
    \For{$q\leftarrow 1$ \KwTo $Q$}{
    $v_q \leftarrow q$-th element of the set $\set{V}$\;\label{alg:lbml-omp_candidateIndexToBeTested}
    $\set{Y}_q=\braces{W_{N-1}\prt{\set{T}^{(i)}}\setminus \braces{g+1}}\cup v_q$\;
    $\v{U}_{q}\leftarrow \v{P}_{\inset{\v{A}_{\set{S}}\prt{\v{\theta}^g}}{\set{Y}_q}}$\;
    }

    $\hat{q} \leftarrow \unset{\opn{argmin}}{q\in [Q]}\left\{\opn{det}\left(\v{U}_{q}\hat{\v{C}}_{\v{x}_{\set{S}}}\v{U}_{q}+\frac{\opn{tr}\prt{\v{U}_{q}^{\perp}\hat{\v{C}}_{\v{x}_{\set{S}}}}}{N-\lvert\set{Y}_q\rvert}\v{U}_{q}^{\perp}\right)\right\}$\;\label{alg:lb-ml-omp_MLDecisionRule}
    $\set{T}^{(i+1)}\leftarrow\set{T}^{(i)}\cup \braces{v_{\hat{q}}}$\;\label{alg:lbml-omp_AddCandToSupportSet}
    $\v{h}^{(i+1)}\leftarrow \unset{\opn{argmin}}{\v{z}\inc{g+1}}\braces{\norm{\v{x}_{\set{D}}-\v{\v{Bz}}}{2}\textrm{, }\supp{z}\subset \set{T}^{(i+1)}}$\;\label{alg:lbml-omp_OrthogonalProjStep}
    }
    $\hat{\v{\theta}}\leftarrow \inset{\v{\theta}^g}{\set{T}^{(K)}\setminus \braces{g+1}}$\;
    \BlankLine
    \caption{Proposed LBML-OMP Algorithm}
    \label{alg:lbml-omp}
    \Out{Estimated DOAs $\hat{\v{\theta}}$}
\end{algorithm}

The asymptotic ML rule for DOA estimation, despite being optimum in the mean-square sense, is a commonly unfeasible approach for most practical scenarios with moderate to large numbers of source signals, if used in the way presented in the literature \cite{VanTrees2002}. The reason for that lies in the requirement of its evaluation over a $D$-dimensional grid search. Instead, the devised approach in (\ref{eq:optProblbml-omp}) requires a one-dimensional evaluation over only $Q$ points.

According to \cite{Dai2009}, the computational complexity of OMP is $\mathcal{O}\prt{K\textrm{DoF}(g+1)}$, under the model in (\ref{eq:CSvecIndexedRecSigCovMat}). The dominant stage is the computation of the pseudoinverse (projection). In the case of LBML-OMP, the whole process is similar, except for the projections $\v{U}_{q}$ (most costly part). Evaluating (\ref{eq:optProblbml-omp}) for each candidate support has about the same complexity as that of one OMP iteration, which leads to a total complexity $Q+1$ times higher than that of OMP, i.e., $\mathcal{O}\prt{(Q+1)K\textrm{DoF}(g+1)}$.

\section{Simulation Results and Discussion}\label{sec:resultsAndDiscussion}

In the simulations, we consider the Uniform Linear Array (ULA),
Two-Level Nested Array (NAQ2), 2nd-Order Super Nested Array (SNAQ2),
Minimum Hole Array (MHA), and Minimum Redundancy Array (MRA)
\cite{Ye2009,Liu2016,Pal2010-2,Moffet1968,Vertatschitsch1986}
schemes. {In the performance evaluation, we employ the Optimal
Sub-Pattern Assignment (OSPA) \cite{Shi2017} metric given by}
\begin{align}
    \begin{split}
        &\textrm{OSPA}(\hat{\v{\theta}},\v{\theta})=\\
        &\sqrt{\frac{1}{LD}\sum_{l=1}^{L}\prt{\unset{\opn{min}}{\set{P}\in\set{P}_D}\sum_{d=1}^{\hat{D}}g^{\prt{\phi}}\prt{[\hat{\v{\theta}}]_{d}^{l},\sqb{\inset{\v{\theta}}{\set{P}}}_{d}}}^2+\phi^2\prt{D-\hat{D}}},
    \end{split}
\end{align}
where $[\hat{\v{\theta}}]_{d}^{l}$ is $d$-th source DOA estimate for
the $l$-th trial, $L$ is the total number of trials, $D$ is the
total number of DOAs to be estimated, $\set{P}_D$ denotes the set of
permutations on $[D]$,
$g^{(\phi)}(\hat{\theta},\theta)=\opn{min}(\phi,|\hat{\theta}-\theta|)$,
and $\phi$ is a penalty parameter for the individual direction
estimate and enumeration bias. {The set of simulation parameters are
shown in Table \ref{tab:param_set}, which contains the number of
sensors, the number of trials for each figure, the normalized
inter-sensor spacing, the grid dimension, the number of candidates
for LBML-OMP, as well as the OSPA penalty parameter. The DOAs assume
the values -0.3426$\pi$, -0.2947$\pi$, -0.2889$\pi$, -0.2820$\pi$,
0.2947$\pi$ radians,} intentionally close to each other to assess
the performance of the algorithms.
\begin{table}
    \caption{Set of Parameters}\label{tab:param_set}
    \centering
    {\begin{tabular}{ccc}
        \hline\hline
        N (sensors) & L (trials) & $\nu/\lambda$ (sensor spacing)\\
        8 & \vtop{\hbox{\strut 5,000 (Fig.~\ref{fig:snaq2RmseSnr}, \ref{fig:lbmlOmpRmseSnr}, \ref{fig:snaq2RmseSnp}, and \ref{fig:error_eta_2})}\hbox{\strut~~~  50,000 (Fig.~\ref{fig:moresources than sensors} and \ref{fig:asymppropetaprime})}} & 1/2\\
        \hline
        g (grid dimension) & Q (number of candidates)  & $\phi$ (OSPA penalty)\\
        1024     & 11 & 0.0430\\
        \hline
    \end{tabular}}
\end{table}
For the case that $\hat{D}>D$,
$\textrm{OSPA}(\hat{\v{\theta}},\v{\theta})=\textrm{OSPA}(\v{\theta},\hat{\v{\theta}})$.
This metric is similar to RMSE, but it allows the assessment of the
error when the number of sources is not correctly estimated, usually
employed in multi-target tracking. The number of candidates was set
to $Q=11$. First, we analyze the behavior of LBML-OMP and then
evaluate the EDCTM model.

\subsection{LBML-OMP}

\begin{figure}[h!]
    \centerline{\includegraphics[width=\columnwidth]{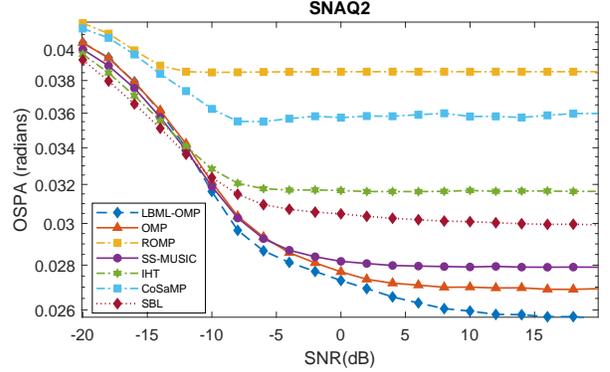}}
    \vspace{-3mm}
    \caption{OSPA against SNR with SNAQ2 geometry for LBML-OMP, OMP, ROMP, SS-MUSIC, IHT, CoSaMP and SBL. $T=50$ snapshots.}
    \label{fig:snaq2RmseSnr}
\end{figure}

Consider the OSPA curves against SNR in Fig.~\ref{fig:snaq2RmseSnr}
using an SNAQ2 arrangement under the DCTM model\footnote{We have
decided to use DCTM instead of EDCTM to evaluate both contributions
of this work individually, although they can obviously benefit from
the synergy of being simultaneously employed.}. Notice that for
almost all the SNR range the devised strategy outperforms classical
CS algorithms like IHT \cite{Blumensath2009}, CoSaMP,
\cite{Needell2009}, and the more recently developed Sparse Bayesian
Learning (SBL) \cite{Gerstoft2016}, as well as coarray dense
approaches like SS-MUSIC. Observe also that LBML-OMP is more costly
than those algorithms, except SBL,  but leads to increased accuracy
due to the selection of candidates following an optimum rule.

At this point, we emphasize that the optimum ML rule in its original version is known to asymptotically achieve the Cramèr-Rao Lower Bound (CRLB) and is asymptotically unbiased (asymptotically efficient DOA estimator) \cite{VanTrees2002,Jaffer1988}. Thus, given an angular sector that comprises the uncertainty region (candidates list) for the OMP response, the derived extended ML rule performs the candidate selection reducing local errors in a feasible way. This rule reduces the potential mismatch between the OMP guess and the right support set element. In this way, it benefits from the synergy between CS-based DOA estimators and asymptotically optimum statistical estimators at the same time.

\begin{figure}[h!]
    \centerline{\includegraphics[width=\columnwidth]{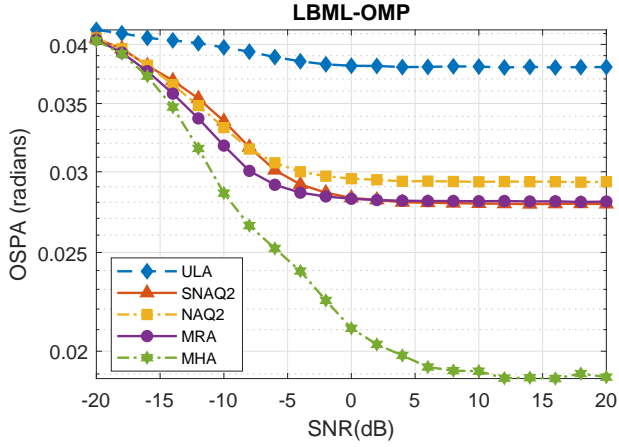}}
    \vspace{-3mm}
    \caption{OSPA against SNR for the proposed LBML-OMP with multiple geometries: ULA, SNAQ2, NAQ2, MRA, and MHA. $T=50$ snapshots.}
    \label{fig:lbmlOmpRmseSnr}
\end{figure}

\begin{figure}[h!]
    \centerline{\includegraphics[width=\columnwidth]{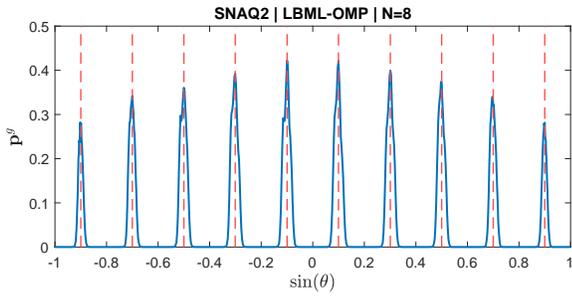}}
    \vspace{-3mm}
    \caption{Identifying more sources than sensors. $N=8$ sensors and $D=10$ sources. $\v{p}^g$ against the sine of the sources DOAs with SNAQ2 geometry for LBML-OMP. Sine of sources DOAs uniformly spaced from -0.9 to 0.9. SNR$=20$ dB. $T=200$ snapshots.}
    \label{fig:moresources than sensors}
\end{figure}

Fig.~\ref{fig:lbmlOmpRmseSnr} shows a comparison among the OSPA curves for LBML-OMP under several geometries: four non-uniform structures (NAQ2, SNAQ2, MHA and MRA) and a ULA. For this case, the arrays are considered to suffer from electromagnetic (EM) coupling. To this end, the B-banded coupling model described in \cite{Liu2016}, given by
    \begin{equation}\label{eq:DoA_DAQ_Coupling}
        \v{x}_\set{S}^{\prime}=\v{G}\v{A}_\set{S}(\v{\theta})\v{s}+\v{n}_{\set{S}},
    \end{equation}
    was employed, where $\v{x}_\set{S}^{\prime}\inc{N}$ is the distorted received signal and $\v{G}\inc{N\tm N}$ is the mutual coupling matrix with elements given by
    \begin{equation}\label{eq:mutualCouplingElements}
        \sqb{\v{G}}_{l,m}=
        \begin{dcases}
            g\prt{\abs{n_l-n_m}} & \text{if $\abs{n_l-n_m}\leq B$}, \\
            0                    & \text{otherwise}.                \\
        \end{dcases}
    \end{equation}
    The coupling coefficients were derived from $g(k)=g(1)\exp(-j(k-1)\pi)/k$, for $2\leq k \leq B$, $B=100$ and $g(1)=0.3\exp\prt{j\pi/3}$. The algorithms do not have any information about the matrix $\v{G}$. In Fig.~\ref{fig:lbmlOmpRmseSnr}, it is clear that MHA is the geometry with best performance for use with LBML-OMP. However, MHA as well as MRA do not have a closed-form solution for the positioning of sensors \cite{VanTrees2002,Liu2016}. Due to that, LBML-OMP can alternatively be employed with SNAQ2, the array with best results following MHA/MRA for most of the SNR range. In addition, it is worth mentioning that ULA exhibits the worst performance, justifying the use of non-uniform arrays for DOA estimation with the proposed approach. In Fig.~\ref{fig:moresources than sensors} we demonstrate the LBML-OMP capability of identifying more sources than sensors ($N=8$ sensors and $D=10$ sources), due to the use of projection matrices that projects the coarray measurements into a full-column rank array manifold even for that condition.

\subsection{EDCTM}

\begin{figure}[h!]
    \centerline{\includegraphics[width=\columnwidth]{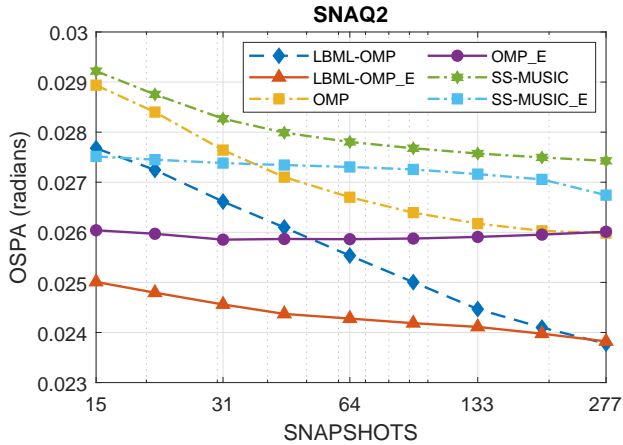}}
    \vspace{-3mm}
    \caption{OSPA against snapshots with SNAQ2 geometry for OMP, LBML-OMP and SS-MUSIC using EDCTM (appended by ``\_E'') and DCTM. SNR$=10$ dB.}
    \label{fig:snaq2RmseSnp}
\end{figure}

In Fig.~\ref{fig:snaq2RmseSnp}, curves named with the suffix ``\_E'' indicate the use of EDCTM, while its absence means that DCTM was employed. Clearly, EDCTM enhances the estimation performance of all algorithms as compared to the performance with the standard DCTM. Notice that EDCTM is also important for theoretical reasons because it might help to improve theoretical results in error analysis of coarray models. The improvement is more prominent for few snapshots and the models converge to the same response after enough data becomes available, according to what was discussed in Section \ref{sec:CsCoarrayDoAEstimation}.

\begin{figure}[h]
    \centerline{\includegraphics[width=\columnwidth]{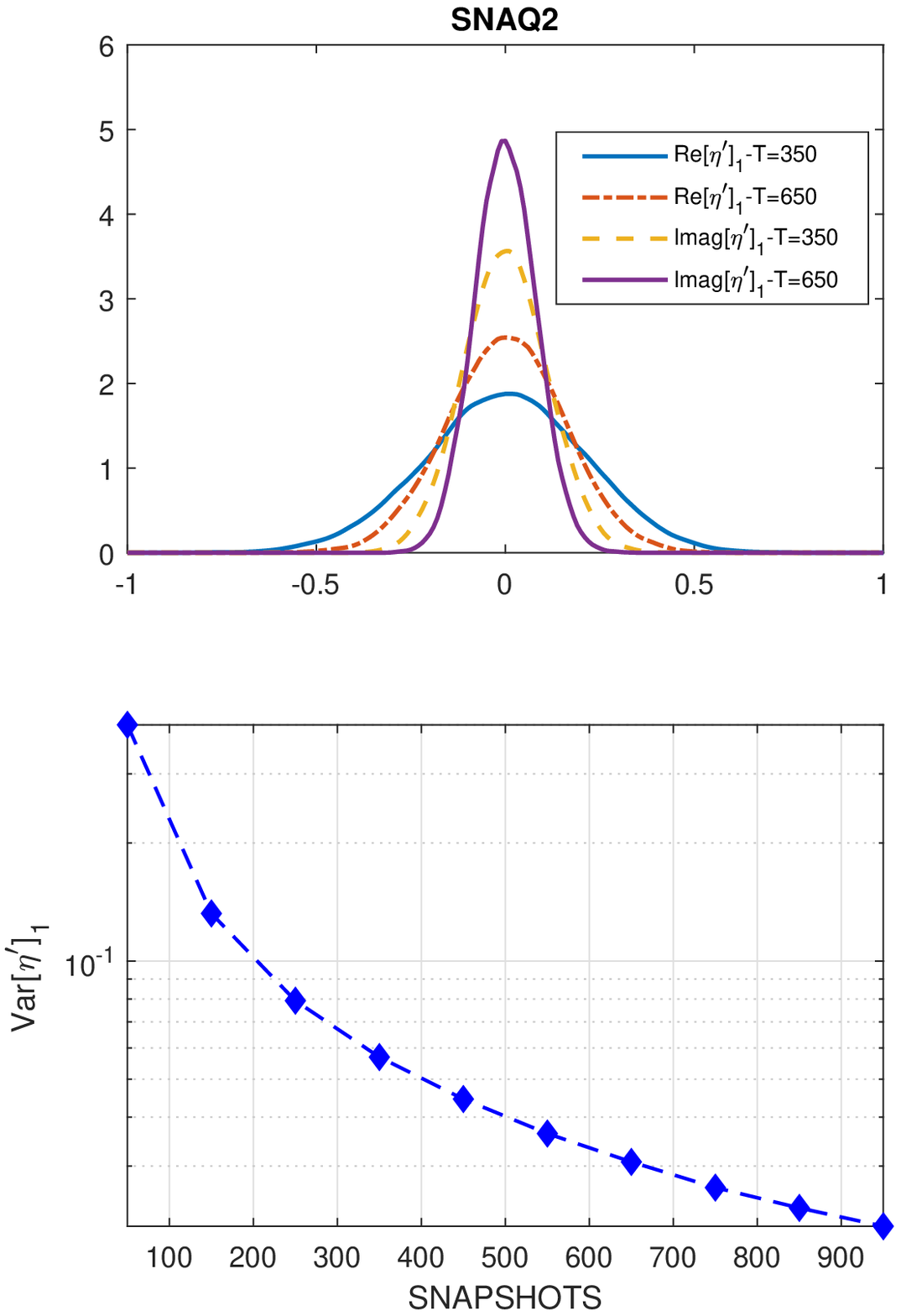}}
    \vspace{-7mm}
    \caption{Statistical and asymptotic properties of $\sqb{\v{\eta}^\prime}_1$. Top: empirical pdf for $\mathfrak{Re}\braces{\sqb{\v{\eta}^\prime}_1}$ and $\mathfrak{Im}\braces{\sqb{\v{\eta}^\prime}_1}$ for $T=350$ and $T=650$ snapshots. Bottom: variance of $\sqb{\v{\eta}^\prime}_1$ against snapshots. SNR$=10$ dB.}
    \label{fig:asymppropetaprime}
\end{figure}

\begin{figure}[h!]
    \centering
    \includegraphics[width=\columnwidth]{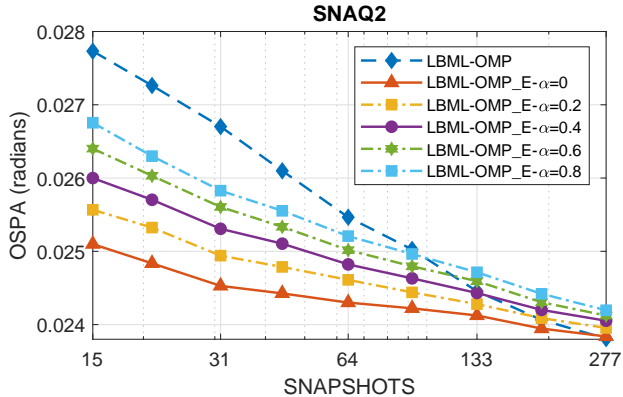}
    \vspace{-3mm}
    \caption{OSPA against snapshots with SNAQ2 geometry for LBML-OMP using DCTM and EDCTM (appended by ``\_E'') for $\alpha\in \braces{0,0.2,0.4,0.6,0.8}$. SNR$=10$ dB.}
    \label{fig:error_eta_2}
\end{figure}

The statistical and asymptotic properties of EDCTM derived in Subsection \ref{sec:asympprop} are illustrated in Fig. \ref{fig:asymppropetaprime}. For that, we have obtained the empirical pdf's for the imaginary and real parts of $\sqb{\v{\eta}^\prime}_1$ for two cases: $T=350$ and $T=650$ snapshots. Moreover, we have also simulated its variance, i.e., $\expt{\abs{\sqb{\v{\eta}^\prime}_1}^2}$. Note that the mean of $\sqb{\v{\eta}^\prime}_1$ is indeed 0 for both imaginary and reals parts, and it does not depend on the number of snapshots. On the contrary, its variance is largely dependent on the amount of available data and tends to zero as we increase the number of snapshots. These conclusions agree with the theoretical analysis conducted in Subsection \ref{sec:asympprop}.
To evaluate the effects of an estimated $\v{\eta}^{\prime}$, denoted by $\hat{\v{\eta}}^{\prime}$, we proceed as follows. Assume that $\hat{\v{\eta}}^{\prime}$ is obtained by an estimator $\Delta$. In this case,
\begin{equation}
    \hat{\v{\eta}}^{\prime}_{\Delta}=\v{\eta}^{\prime}+\Tilde{\v{\eta}}^{\prime},
\end{equation}
where $\Tilde{\v{\eta}}^{\prime}$ is an additive zero-mean white circular complex Gaussian distributed random vector\footnote{Note that the entries of $\v{\eta}^\prime$ are not Gaussian since they correspond to a complex weighted sum of the off-diagonal entries of a complex Wishart distributed matrix. However, by resorting to the central limit theorem and its variants, the Gaussian approximation is a reasonable assumption. This can be empirically verified in Fig. \ref{fig:asymppropetaprime} (top).} (estimation error) with covariance matrix $\v{C}_{\Tilde{\v{\eta}}^{\prime}}=\expt{\Tilde{\v{\eta}}^{\prime}\Tilde{\v{\eta}}^{\prime H}}=\sigma_{\Tilde{\v{\eta}}^{\prime}}^2\v{I}$. Moreover, consider that $\sigma_{\Tilde{\v{\eta}}^{\prime}}^2 = \alpha\norm{\v{\eta}^{\prime}}{2}^2/\card{D}$, where $\alpha\in \braces{0,0.2,0.4,0.6,0.8}$ controls the quality of the estimation process. In order to evaluate the effects of the estimation error $\Tilde{\v{\eta}}^{\prime}$ for the DOA estimates, consider what is shown in Fig. \ref{fig:error_eta_2}. Notice that $\alpha=0$ corresponds to the \emph{a priori} knowledge of $\v{\eta}^{\prime}$. Indeed, there is a substantial gain in performance even for $\alpha=0.6$ (poor estimation) in the small-sample region.

\section{Conclusion}\label{sec:conclusions}

We have developed an enhanced coarray estimation model termed EDCTM along with the LBML-OMP algorithm, which exploits the correlation-maximization in OMP and the asymptotic ML DOA estimator. The results show that the proposed approaches outperform existing techniques for DOA estimation and highlight errors introduced by the traditional difference coarray transformation in finite-sample scenarios.

\bibliographystyle{IEEEtran}
\bibliography{mybib}

\end{document}